\documentclass[preprint, 12pt, a4paper]{elsarticle}

\usepackage{amssymb}

\usepackage{float}
\usepackage{url}
\usepackage{color}
\usepackage{cancel}

\restylefloat{table}

\journal{SoftwareX}

\begin{document}

\begin{frontmatter}



\title{MateriApps LIVE! and MateriApps Installer: Environment for starting and scaling up materials science simulations}

\author[label1]{Yuichi Motoyama}
\author[label1]{Kazuyoshi Yoshimi}
\author[label1]{Takeo Kato}
\author[label2,label3,label1]{Synge Todo}

\address[label1]{Institute for Solid State Physics, The University of Tokyo, Chiba 277-8581, Japan}
\address[label2]{Department of Physics, The University of Tokyo, Tokyo  113-0033, Japan}
\address[label3]{Institute for Physics of Intelligence, The University of Tokyo, Tokyo 113-0033, Japan}

\begin{abstract}
In our current era, numerical simulations have become indispensable theoretical and experimental tools for use in daily research activities, particularly in the materials science fields. However, the installation processes for such simulations frequently become problematic because they depend strongly on the device environment, and troubleshooting those processes is a challenging task for beginners. To minimize such difficulties, we created MateriApps LIVE! and MateriApps Installer, which can solve most of the related issues. Specifically, MateriApps LIVE! offers a virtual environment in which users can quickly try out computational materials science simulations on a personal computer  while MateriApps Installer provides a comprehensive set of shell scripts for use when installing software on Unix, Linux, macOS, and supercomputer systems. Herein, we provide detailed descriptions of MateriApps LIVE! and MateriApps Installer together with illustrative examples of their use.
\end{abstract}


\end{frontmatter}




\begin{table}[H]
\begin{tabular}{|l|p{6.5cm}|p{8.0cm}|}
\hline
C1 & Current code version & 3.3 (MateriApps LIVE!), 
1.1.0 (MateriApps Installer) \\
\hline
C2 & Permanent link to code/repository used for this code version &
https://sourceforge.net/projects/materiappslive (MateriApps LIVE!), https://github.com/wistaria/MateriAppsInstaller (MateriApps Installer)\\
\hline
C3 & Code Ocean compute capsule & \\
\hline
C4 & Legal Code License   & GPL v3 \\
\hline
C5 & Code versioning system used & git \\
\hline
C6 & Software code languages, tools, and services used & shell script \\
\hline
C7 & Compilation requirements, operating environments \& dependencies & Unix, Linux, macOS, Windows \\
\hline
C8 & If available Link to developer documentation/manual & \\
\hline
C9 & Support email for questions & mainstaller-dev@issp.u-tokyo.ac.jp \\
\hline
\end{tabular}
\caption{Code metadata}
\label{tab:metadata} 
\end{table}



\section{Motivation and Significance} 
\label{sec:motivation}

Numerical simulations are now indispensable tools for both theorists and experimentalists in materials science
 because precise simulations can help elucidate mechanisms behind novel phenomena, predict the various physical properties of new materials, and help identify candidate materials with desired performance characteristics.
In computational materials science, new methods are being proposed daily and a wide range of software is being developed actively. 
Now, because of diverse cutting-edge software developments, users can easily employ highly parallelized numerical simulations based on efficient algorithms and state-of-the-art functionalities without the necessity of writing code by themselves. 
In an effort to consolidate and disseminate information on materials science software, we launched the MateriApps portal site for material science simulations in 2013~\cite{MateriApps, Konishi15}, thus providing a well-organized forum where users could quickly search and find materials science software with desired functionalities, as well as obtain detailed information on each application.

However, while the promotion of such software dissemination activities requires preparing environments in which users can quickly start testing materials science software, the process of installing such software frequently becomes problematic. This is because most such software requires the use of a command-line interface on the terminal window, which is an extremely high barrier for beginners to overcome. Furthermore, even during hands-on software sessions in which instructors teach software installation processes in detail, such procedures are still troublesome because they depend on the participant's personal computer (PC) environment. Nevertheless, it is clear that most such issues could be resolved by utilizing an environment in which users can easily install software without concerning themselves with operating system (OS) and software details. For this purpose, we developed MateriApps LIVE!~\cite{MateriAppsLive}, which offers a virtual environment in which users can quickly try out computational materials science simulation on PCs. 

Suppose users want to execute simulations for real problems that require significant system resources. In such cases, virtual environments are not practical because the memory size is minimal, and there is a non-negligible virtualization overhead. In addition, there are numerous situations where virtual environments are unavailable due to machine architectures, company regulations, computer system operating policies, and other issues. In those cases, users must install and maintain the software by themselves while also making the appropriate setting changes to the compiler options and numerical libraries that will leverage the computer system's performance. 
Another critical issue is the need to manage the installation scripts consistently and organize them appropriately for reuse. This is particularly important when preinstalling software on large-scale supercomputers. To overcome these problems, we have developed MateriApps Installer~\cite{MateriAppsInstaller},  which provides a comprehensive set of shell scripts for software installations on Debian GNU/Linux (Ubuntu), CentOS (RedHat), macOS, and many supercomputer system environments.

\begin{figure}
    \centering
     \includegraphics[width= 0.9\linewidth]{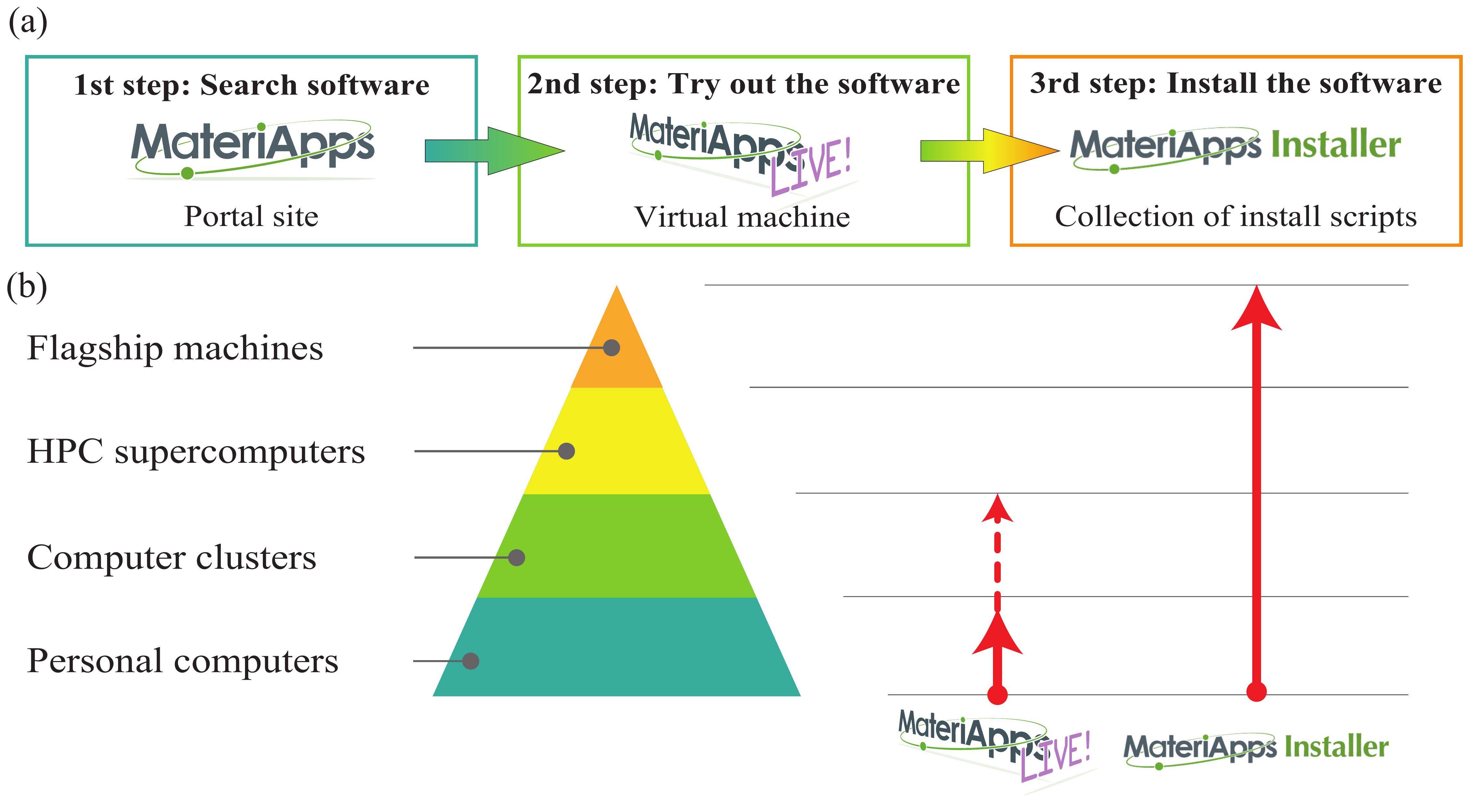}
    \caption{(a) Three-step flow diagram, indicating the relationships among MateriApps, MateriApps LIVE!, and MateriApps Installer.
    (b) Schematic of computing environments supported by MateriApps LIVE! and MateriApps Installer.}
    \label{fig:schematicfigs}
\end{figure}

We summarize common usage images of MateriApps, MateriApps LIVE!, and Materials Installer in Fig.~\ref{fig:schematicfigs}~(a). First, users utilize the MateriApps portal site to find the suitable software for their purposes.
Next, they try software on their PCs using MateriApps LIVE!.
Finally, if they hope to maximize software's performance, they start to use MateriApps Installer.
The target computers that are assumed in MateriApps LIVE! and MateriApps Installer are also summarized in Fig.~\ref{fig:schematicfigs}~(b).

The remainder of this paper is organized as follows: In Sec.~\ref{sec:description}, we provide a detailed explanation of the architecture and functionalities of MateriApps LIVE! and MateriApps Installer, while in Sec.~\ref{sec:example}, we give illustrative examples of how to use both products. In Sec.~\ref{sec:impact}, we discuss the effects of MateriApps LIVE! and MateriApps Installer on the promotion of materials science simulations, while in Sec.~\ref{sec:conclusion}, we provide concluding remarks.

\section{Software Description}
\label{sec:description}

\subsection{Software Architecture}
\label{sec:architecture}

MateriApps LIVE!, which is a Debian GNU/Linux OS that includes various preinstalled materials science software packages, is distributed as a virtual machine snapshot in the open virtualization appliance (OVA) format. As a first step, each simulation software package is prepared as a Debian package and uploaded to the MateriApps package repository, where it is managed with Debian's advanced packaging tool (APT). Later, as each original software version is upgraded, its corresponding Debian package is updated and uploaded accordingly so that users can promptly use the latest version of the simulation software. The MateriApps LIVE! OVA file is generated automatically using the Packer image-building tool. Note that Debian packages are freely available from the MateriApps package repository and can be installed, via APT, on systems already running on Debian GNU/Linux (bullseye/buster/stretch) and Ubuntu (focal/bionic).

Separately, MateriApps Installer is a collection of shell scripts (compatible with the POSIX standard) that is intended for use when installing materials science programs.
As a matter of design policy, MateriApps Installer limits its requirements to just a few fundamental tools such as POSIX-compatible shells, \verb|make|, and \verb|tar|.
However, in addition to materials science programs, MateriApps Installer offers install scripts for dependencies such as GCC and CMake, even though the official or standard package managers (e.g., Debian packages, RedHat RPM, and Homebrew) are normally recommended.
In use, MateriApps Installer installs each version of each tool/application into a separate directory.
For example, in the case of CMake 3.22.3, all files are installed into
\begin{verbatim}
$MA_ROOT/cmake/cmake-3.22.3-1
\end{verbatim}
so they do not interfere with the previous versions and other tools.
The last number added, \verb|1|, denotes the MateriApps Installer script revision (e.g., the set of compile options).
Note that users can choose a different installation location for each software package by changing the \verb|$MA_ROOT| environment variable.
Additionally, since MateriApps Installer installs shell scripts for setting environment variables, such as \verb|$PATH|, for each tool/application/version, users can activate each software package as needed.
When designing MateriApps Installer, we separated the installation procedure from the linking process, so new software installations do not affect anything else until the linking is complete.
Additionally, the software version is maintained as closely as possible to MateriApps LIVE!, and for ease of maintenance, MateriApps LIVE! and MateriApps Installer utilize the same source code patches.

\subsection{Software Functionalities}
\label{sec:functionality}

MateriApps LIVE! offers an environment in which users can ``test drive'' computational materials science simulations freely using user's PC or similar device.
The entire environment required to launch the tutorials, including the OS (Debian GNU/Linux), editors, materials science applications, and visualization tools, is provided as an OVA file.
After the OVA file is imported into a virtual machine software product such as VirtualBox, various computational materials science simulators immediately become available.  

On the other hand, MateriApps Installer is a collection of shell scripts designed to assist users when installing computational material science applications that includes all the necessary tools and libraries for various computing environments.
Currently, MateriApps Installer supports Debian GNU/Linux (Ubuntu), CentOS (RedHat), macOS, and many major supercomputer systems.
In Table~\ref{table:list_malive}, we summarize the software packages and tools that are currently preinstalled in MateriApps LIVE! and supported by MateriApps Installer (in bold).
Here, we marked $*$ to 2dmat in the table since 2dmat is supported by MateriApps Installer but not preinstalled in MateriApps LIVE!.
This is because 2dmat is the software using MPI parallelization for large size of computations and thus it is expected that efficient calculations are not performed on MateriApps LIVE!.

\begin{table}[H]
\begin{tabular}{|p{3 cm}|p{10 cm}|}
\hline
Type & Software names\\
\hline
Applications & AkaiKKR~\cite{AkaiKKR,Akai1982}, {\bf ALPS}~\cite{ALPS,Alet2005,Albuquerque2007,Bauer2011}, CASINO (only setup tool)~\cite{CASINO,Needs2020}, CONQUEST~\cite{CONQUEST,Bowler2006}, {\bf DCore}~\cite{DCore,Shinaoka2021}, DDMRG~\cite{DDMRG}, {\bf DSQSS}~\cite{DSQSS, MOTOYAMA2021107944}, GAMESS (only setup tool)~\cite{GAMESS,Barca2020}, Gromacs~\cite{Gromacs,Pronk2013,Abraham2015,Pall2015}, {\bf HPhi}~\cite{HPhi, KAWAMURA2017180}, {\bf LAMMPS}~\cite{LAMMPS,Thompson2022}, {\bf mVMC}~\cite{mVMC, MISAWA2019447}, OCTA~\cite{OCTA}, {\bf OpenMX}~\cite{OpenMX,Ozaki2003,Ozaki2004}, {\bf Quantum ESPRESSO}~\cite{QUANTUMESPRESSO,Giannozzi2009,Giannozzi2017}, {\bf RESPACK}~\cite{RESPACK, NAKAMURA2021107781}, SALMON~\cite{SALMON,Noda2019}, SMASH~\cite{SMASH}, {\bf TeNeS}~\cite{TeNeS, motoyama2021tenes},  {\bf TRIQS/CTHYB}~\cite{CTHYB,Seth2016}, {\bf TRIQS/DFT tools}~\cite{DFTTools, Aichhorn2016}, xTAPP~\cite{xTAPP}, {\bf 2dmat}$^*$~\cite{2dmat, 2dmat-paper}\\
\hline
Visualization / modeling tools & C-Tools~\cite{CTOOLS}, CIF2Cell~\cite{CIF2CELL,Torbjorn2011}, FermiSurfer~\cite{FERMISURFER, KAWAMURA2019197}, gnuplot, Open Babel~\cite{OPENBABEL,Boyle2011}, OpenDX~\cite{OPENDX}, OVITO~\cite{OVITO,Alexander2010}, ParaView~\cite{ParaView}, Pymol~\cite{Pymol,PyMOLcitation}, Rasmol~\cite{Rasmol,Sayle1995}, Tapioca~\cite{TAPIOCA}, VESTA~\cite{VESTA,Momma2011}, VMD (only setup tool)~\cite{VMD,Humphrey1996}, XCrySDen~\cite{XCRYSDEN,Kokalj1999}, Wannier90~\cite{Wannier90,Pizzi2020}\\
\hline
Data analysis / supplementary tools & ALAMODE~\cite{ALAMODE,Tadano2014}, {\bf ALPSCore}~\cite{ALPSCORE,Wallerberger2018}, BSA~\cite{BSA}, {\bf PHYSBO}~\cite{PHYSBO, motoyama2021bayesian} \\
\hline
Editors & Mousepad, Nano, Vim, Emacs \\
\hline
Development tools &  {\bf GCC (gcc, g++, gfortran)}, {\bf CMake}, {\bf Git}, {\bf Python}, etc \\
\hline
Libraries & BLAS (OpenBLAS), {\bf Boost}, {\bf Eigen3}, {\bf HDF5}, {\bf LAPACK}, etc \\
\hline
\end{tabular}
\caption{List of software packages and tools preinstalled in MateriApps LIVE! and supported by MateriApps Installer (in bold).
Here, 2dmat is marked with $*$ because it is supported by MateriApps Installer but not preinstalled on MateriApps LIVE!.}
\label{table:list_malive}
\end{table}

\section{Illustrative Examples}
\label{sec:example}

\subsection{MateriApps LIVE!} \label{SubSec:MALIVE}

\subsubsection{Starting MateriApps LIVE!}

MateriApps LIVE! is distributed as a snapshot image of the virtual machine (OVA file) and can be booted within virtualization software such as VirtualBox.
The MateriApps LIVE! distribution package can be downloaded from the official site \url{https://cmsi.github.io/MateriAppsLive/} and can be launched by installing the downloaded package into the virtualization software.
For example, if a user uses VirtualBox, MateriApps LIVE! can be installed by double-clicking the icon of the distribution package.

\begin{figure}
    \centering
    \includegraphics[width=\linewidth]{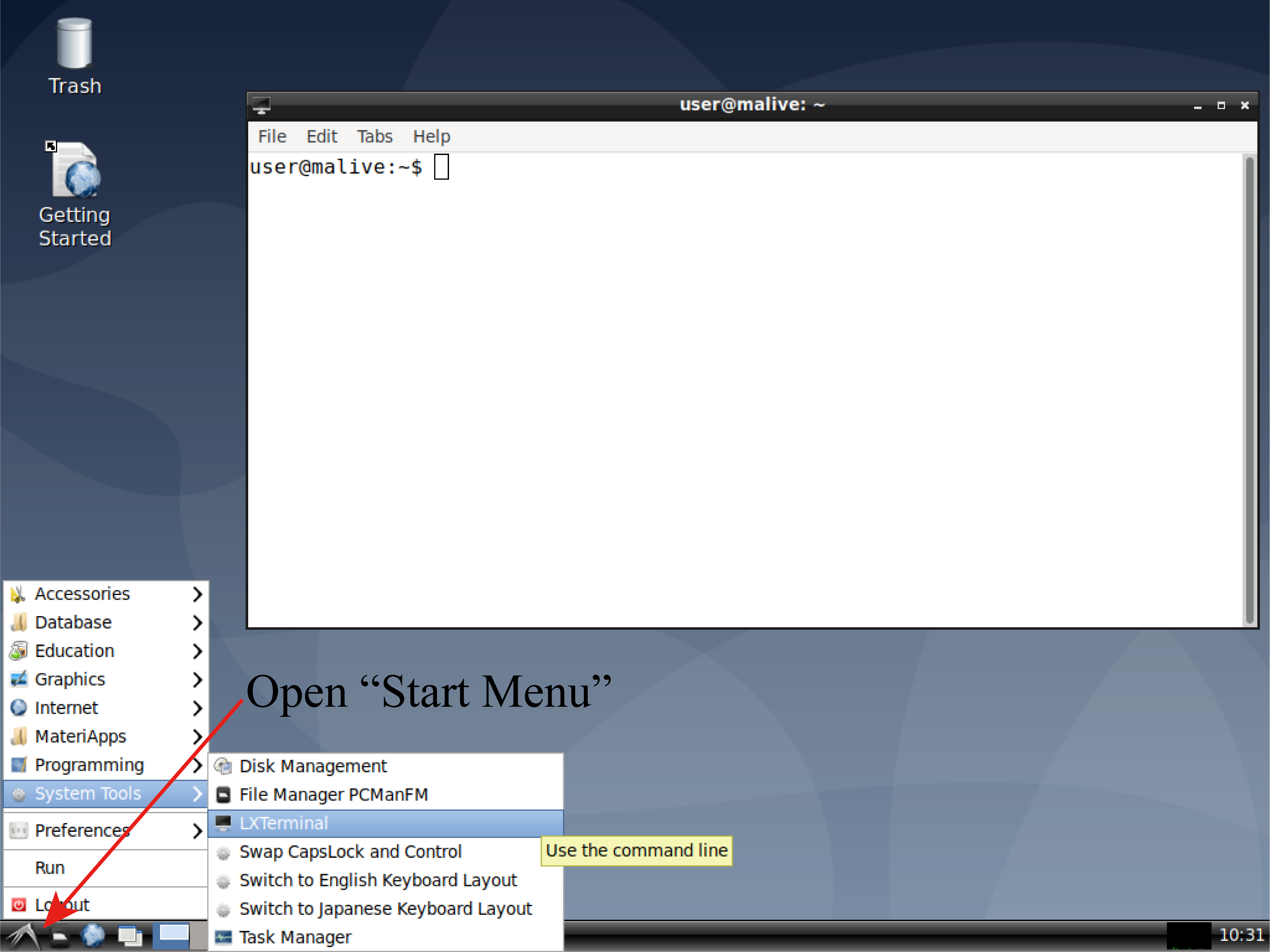}
    \caption{How to start LXTerminal. Open ``Start Menu'' by clicking the leftmost bottom button, and then click ``LXTerminal'' in ``System Tools''.}
    \label{fig:LXTerminal}
\end{figure}

In MateriApps LIVE!, a user with the username ``user'' is predefined with the password ``live''.
After logging into MateriApps LIVE!, the user opens \texttt{LXTerminal} from the ``Start Menu'' (the leftmost icon on the menu bar), then navigates $\to$ System Tools, and then $\to$ LXTerminal (Figure~\ref{fig:LXTerminal}).
This shows that it is possible to execute various software packages in the terminal window without a cumbersome installation process or worrying about environment variable setting.
The executable binary files of the software packages are located in \texttt{/usr/bin}, and the utility tools and example files are installed in \texttt{/usr/share/}{\textit{software}\_name} directory.

\subsubsection{Updating software}

Numerous software packages are being actively developed, and many are being frequently updated.
In the MateriApps LIVE!\ development system, Debian packages are continuously updated according to their upstream upgrades, which means users can easily update their application packages via the following steps:\footnote{This procedure updates all the packages in MateriApps LIVE!. If a user wants to update a specific package, e.g., $\mathcal{H}\Phi$, he or she should use \texttt{sudo apt install hphi} in the second step.}
\begin{verbatim}
$ sudo apt update
$ sudo apt upgrade
\end{verbatim}

\subsubsection{Running a simulation}

As an example, we will now demonstrate how to perform the exact diagonalization for quantum lattice models by using the $\mathcal{H}\Phi$\cite{KAWAMURA2017180} software package.
Let us start by showing how to calculate the temperature dependence of the magnetic susceptibility of the antiferromagnetic Heisenberg (AFH) chain using the thermal pure quantum (TPQ) method~\cite{Sugiura2012,Sugiura2013}.

\begin{figure}
    \centering
    \includegraphics[width=\linewidth]{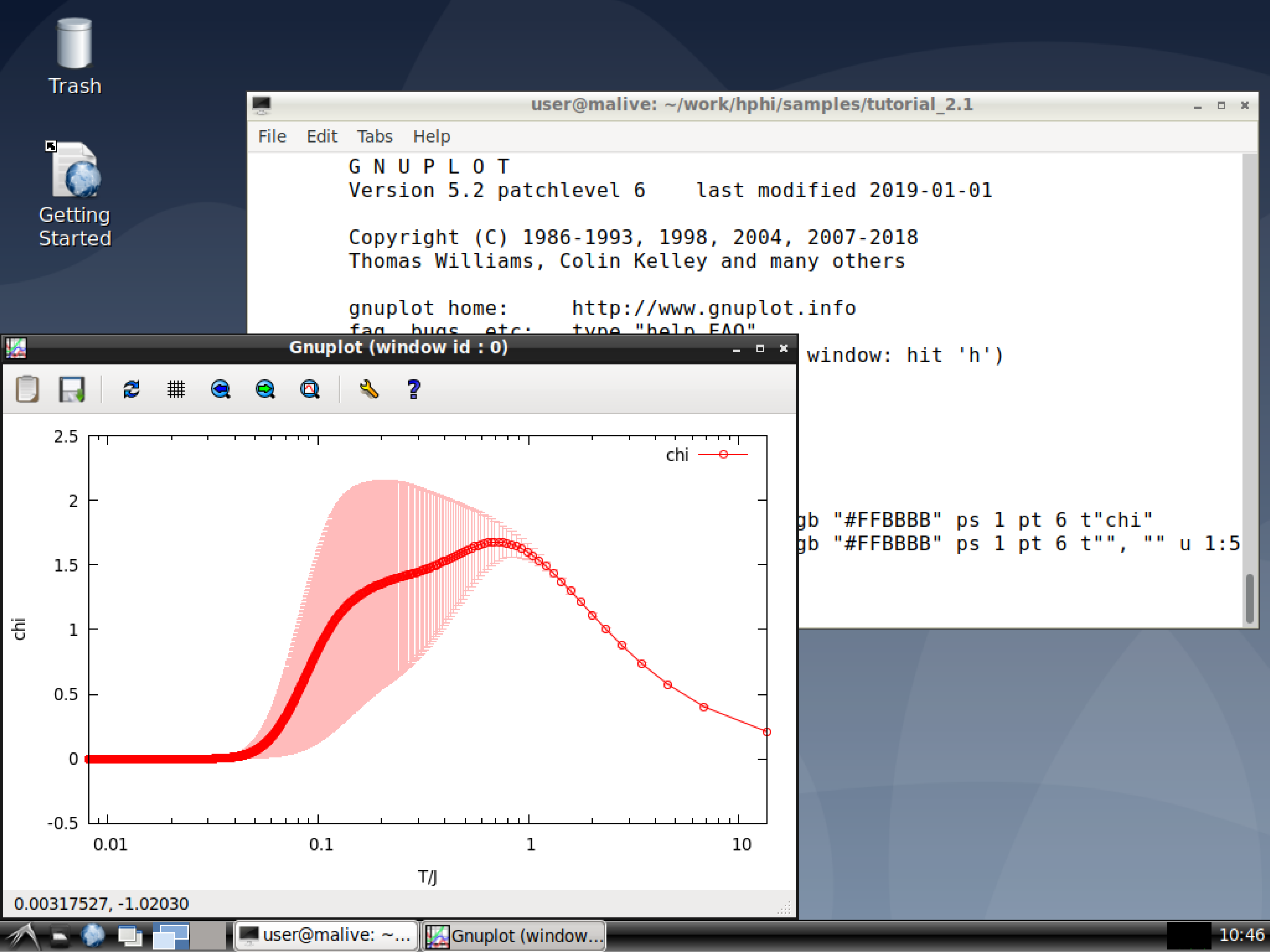}
    \caption{Magnetic susceptibility of an antiferromagnetic Heisenberg chain calculated via $\mathcal{H}\Phi$ running on MateriApps LIVE! using the TPQ method.}
    \label{fig:TPQ}
\end{figure}

First, we copy the sample files from \texttt{/usr/share/hphi/samples} to a working directory, e.g., \texttt{\$HOME/hphi}:
\begin{verbatim}
$ cd
$ mkdir hphi && cd hphi
$ cp -r /usr/share/hphi/samples ./
\end{verbatim}
In $\mathcal{H}\Phi$ v3.5.0, which is currently the latest version, users can find the sample input file for the TPQ calculation of the AFH chain at \texttt{tutorial\_2.1/stan3.in}.
At that point, the user executes $\mathcal{H}\Phi$ using the following commands:
\begin{verbatim}
$ cd $HOME/hphi/samples/tutorial_2.1
$ HPhi -s stan3.in
\end{verbatim}
The results are then saved into the \texttt{output} directory.
By using a postprocessing script \texttt{AveFlct.py} to calculate susceptibility, users can generate the magnetic susceptibility as follows:
\begin{verbatim}
$ python3 AveFlct.py
\end{verbatim}
The generated data are then saved into a file named \texttt{ave\_Flct.dat}.
Finally, the result is visualized by using Gnuplot (Fig.~\ref{fig:TPQ}):
\begin{verbatim}
$ gnuplot
gnuplot> set log x
gnuplot> set colors classic
gnuplot> set xlabel "T/J"
gnuplot> set ylabel "chi"
gnuplot> pl "ave_Flct.dat" u 1:5:6 w ye lc rgb "#FFBBBB" pt 6 t"", \
"" u 1:5 w lp lt 1 pt 6 t"chi
\end{verbatim}
For $\mathcal{H}\Phi$ details, please refer to the official tutorial~\cite{HPhi}.

\subsection{MateriApps Installer}

In this section, we describe how to set up MateriApps Installer and install tools and applications on Linux OS.
A MateriApps Installer zip file can be downloaded from the official website~\cite{MateriAppsInstaller}.
After opening the zip file, users can obtain all the shell scripts included in MateriApps Installer.
If Git is available, MateriApps Installer can be cloned by typing the following command:
\begin{verbatim}
$ git clone https://github.com/wistaria/MateriAppsInstaller
\end{verbatim}
The MateriApps Installer directory structure is as follows:

\begin{verbatim}
|- apps
|- docs
|- macos
|- scripts
|- setup
|- tools
|- README.md
|- check_prefix.sh
\end{verbatim}
The following scripts and file directories are available.
\begin{itemize}
    \item \verb|check_prefix.sh|: Displays variables that are commonly used in each script, such as the top installation directory.
    \item \verb|docs| directory: A directory containing the manual and its source code.
    \item \verb|macos| directory: A directory containing scripts used to install the necessary tools using Macports.
    \item \verb|setup| directory: A directory containing scripts used to prepare for software installation.
\end{itemize}
The directory structure in \verb|setup|, \verb|tools|, and \verb|apps| is given as follows:
\begin{verbatim}
-- software_name
      |- README.md
      |- download.sh
      |- link.sh
      |- setup.sh
      |- version.sh
      |- install.sh
      |- patch
      |- config
\end{verbatim}
Each file and directory is described as shown below:
\begin{itemize}
    \item \verb|README.md|: Includes a brief introduction of the software and the URLs of the official website.
    \item \verb|download.sh|*: Download the source code archive.
    \item \verb|install.sh|*: Build and install the software.
    \item \verb|link.sh |*: Create symbolic links to installed directories and configuration files.
    \item \verb|setup.sh|*: Extract the prepared source code archive and apply the patch (if it exists).
    \item \verb|version.sh|*: Specify the version to download.
    \item \verb|patch| directory: The directory where the patches are stored.
    \item \verb|config| directory: Additional settings for installation other than the default settings, such as when using the Intel Compiler.
\end{itemize}
Here, a file marked with an * indicates a file that must always be present in the directory. 
Below, we show how to install the tools and software through a step-by-step demonstration.

\subsubsection{Installing tools}

First, perform the initial setup by entering the MateriApps Installer source directory and executing the following command:
\begin{verbatim}
$ sh setup/setup.sh
\end{verbatim}
This command prepares the directory where tools and applications will be installed.
The default location is \verb|$HOME/materiapps/|.\footnote{The install location can be specified by adding \texttt{MA\_ROOT=}\textit{directory}\_\textit{name} in \texttt{\$HOME/.mainstaller}.}
Next, the compiler is installed.
Here, it should be noted that while the compiler installed in the system can also be used for installing other tools and applications, some applications may not compile well with the old version of GCC.
If users have little experience in solving compilation problems, it is strongly recommended that they use the GCC installed by MateriApps Installer.
To install GCC 10, execute the following commands:
\begin{verbatim}
$ cd tools/gcc10
$ sh install.sh
\end{verbatim}
After finishing the installation, execute
\begin{verbatim}
$ sh link.sh
\end{verbatim}
to set up an appropriate symbolic link.

Many recent tools and applications, including the major Linux distributions, include CMake to facilitate installation processes.
However, to avoid potential installation problems, it is recommended that
the latest CMake version is installed by MateriApps Installer.
To install CMake, execute the following commands in the MateriApps Installer source directory:
\begin{verbatim}
$ cd tools/cmake
$ sh install.sh
$ sh link.sh
\end{verbatim}
The other tools can be installed using the same process.
Note that GCC, Git, libffi, and Python3 must be installed in that order before the other tools can be installed.
After installing the above tools, the user should install the remaining tools required for applications, e.g., in alphabetical order (e.g., Boost, Eigen3, FFTW, GSL, HDF5, LAPACK, OpenMPI, OpenSSL, ScaLAPACK, TclTk, and Zlib).

\subsubsection{Installing applications}

After installing the tools, users can install the application by entering the directory of the application to be installed and executing the installation script \verb|install.sh| as follows:
\begin{verbatim}
$ sh install.sh
\end{verbatim}
Users can then check whether the installation was successful by running the check script using the following command:
\begin{verbatim}
$ sh runtest.sh
\end{verbatim}

It should also be noted that if the application installation was terminated during the process, some files may remain in incomplete form in \texttt{\$HOME/materiapps/}\textit{application}\_\textit{name}| directory and will cause errors when reinstallation is attempted.
In such cases, users have to remove the directory of the target application by hand.
If errors are encountered during installation, please read the error messages carefully.
Most errors result when the required tools are not installed.

Once the test has been completed, execute the following command to prepare an appropriate link to the application setting script:
\begin{verbatim}
$ sh link.sh
\end{verbatim}

\subsubsection{Running applications}

Before running the application, the environment variables must be set.
The following command sets the variables required for $\mathcal{H}\Phi$: 
\footnote{For other applications, use \texttt{\$HOME/materiapps/}\textit{application}\_name\texttt{/}\textit{application}\_name\texttt{vars.sh} instead.}
\begin{verbatim}
$ source $HOME/materiapps/hphi/hphivars.sh
\end{verbatim}
The sample files can be found in \texttt{\$HPHI\_ROOT/samples}.
Users can execute $\mathcal{H}\Phi$ using the same process described in Sec.~\ref{SubSec:MALIVE}:
\begin{verbatim}
$ cd
$ mkdir hphi && cd hphi
$ cp -r $HPHI_ROOT/samples ./
$ cd $HOME/hphi/samples/tutorial_2.1
$ HPhi -s stan3.in
\end{verbatim}

\section{Impact}
\label{sec:impact}

MateriApps LIVE! allows computer simulation non-experts to start a simulation in just a few minutes, while MateriApps Installer makes it easy to run simulations on computer clusters and supercomputers, thereby facilitating larger-scale and more accurate simulations than ever before.
As a result, MateriApps LIVE! and MateriApps Installer free users from cumbersome simulation software installation and upgrade processes, thereby enabling them to seamlessly use the same version of the software on any device from users' PCs to state-of-the-art supercomputers (see also Fig.~\ref{fig:schematicfigs}~(b)).
These significant advantages, which cannot be achieved by previous approaches, are expected to contribute significantly to the promotion of computational simulation research by encouraging an increasing number of experimental and industrial researchers to engage in new simulations.

The use of MateriApps LIVE! and MateriApps Installer for materials science simulations has been progressing steadily.
In fact, MateriApps LIVE! has been downloaded more than 10,000 times since its first release in 2013 and won Open Source Excellence Award from SourceForge in 2022.
Additionally, the MateriApps LIVE! environment has been used in more than 30 lectures and workshops attended by more than 1,000 participants.
It has also been utilized in internal corporate training, hands-on seminars, and in universities for undergraduate and graduate student classes. 

Separately, the use of MateriApps Installer has been gradually increasing since the release of version 1.0.0 in March 2021, and it has already been used for installing materials science software on supercomputers at the Institute for Solid State Physics (ISSP) at the University of Tokyo, and the High-Performance Computing Infrastructure (HPCI) in Japan.
We anticipate that other companies will soon provide support, simulation services, and consulting for computational materials science simulations based on the MateriApps environment.

\section{Conclusions}
\label{sec:conclusion}

In this paper, we showed how MateriApps LIVE! and MateriApps Installer help users get started with materials science simulations.
More specifically, we showed that MateriApps LIVE! provides a virtual environment that immediately enables users to try out materials science simulations on users' PC.
This environment dramatically reduces the barriers encountered by non-experts who want to start utilizing various simulation packages in unified OS settings.
Separately, MateriApps Installer provides a comprehensive set of shell scripts to facilitate the installation of software on Unix, Linux, macOS, and supercomputer systems.
Conducting installations via this script collection enables users to maximize the performance of high-end workstations, laboratory-level computer clusters, and supercomputers in materials science simulations.
Since our ultimate goal is promoting further research and development in computational materials science, we believe that it is crucial for us to continue the development of MateriApps LIVE! and MateriApps Installer, as well as the MateriApps portal site, by adding new software packages and updating the currently included software products.

\section*{Acknowledgements}
\label{sec:acknowledgement}

MateriApps Installer was developed under the support of the ``Project for advancement of software usability in materials science'' (PASUMS) in the fiscal year 2020 by the Institute for Solid State Physics (ISSP) at the University of Tokyo.
MateriApps is supported by the ISSP, the University of Tokyo, and the Elements Strategy Initiative Center for Magnetic Materials (ESICMM).
KY was supported by the Japan Society for the Promotion of Science (JSPS) KAKENHI Grant No. 19K03649.

\bibliography{mainstaller}
\end{document}